\documentclass[12pt]{article}%
\usepackage[dvipsnames]{xcolor}
\usepackage{amsfonts}
\usepackage{amsmath}
\usepackage{amssymb}
\usepackage{float}
\usepackage{graphicx}%
\providecommand{\U}[1]{\protect\rule{.1in}{.1in}}
\begin{document}

\title{Supersymmetric solitons and a degeneracy of solutions in AdS/CFT}
\author{Andr\'{e}s Anabal\'{o}n$^{(1)}$ and Simon F. Ross$^{(2)}$\\\smallskip\\$^{(1)}$Departamento de Ciencias, Facultad de Artes Liberales, \\Universidad Adolfo Ib\'{a}\~{n}ez, Avenida Padre Hurtado 750, Vi\~{n}a del
Mar, Chile. \\\smallskip\\$^{(2)}$Centre for Particle Theory, Department of Mathematical Sciences, \\Durham University, South Road, Durham DH1 3LE, U.K.}
\maketitle

\begin{abstract}
We study Lorentzian supersymmetric configurations in $D=4$ and $D=5$ gauged $\mathcal{N}=2$ supergravity. We show that there are smooth $1/2$ BPS solutions which are asymptotically AdS$_{4}$ and AdS$_{5}$ with a planar
boundary, a compact spacelike direction and with a Wilson line on that circle.
There are solitons where the $S^{1}$ shrinks smoothly to zero in the interior,
with a magnetic flux through the circle determined by the Wilson line, which
are AdS analogues of the Melvin fluxtube. There is also a solution with a
constant gauge field, which is pure AdS. Both solutions preserve half of the
supersymmetries at a special value of the Wilson line. There is a phase
transition between these two saddle-points as a function of the Wilson line
precisely at the supersymmetric point. Thus, the supersymmetric solutions are
degenerate, at least at the supergravity level. We extend this discussion to
one of the Romans solutions in four dimensions when the Euclidean boundary is
$S^{1}\times\Sigma_{g}$ where $\Sigma_{g}$ is a Riemann surface with genus $g
> 0$. We speculate that the supersymmetric state of the CFT on the boundary is
dual to a superposition of the two degenerate geometries.

\end{abstract}

\section{Introduction}

There is a long-standing interest in smooth soliton solutions in gravity.
There are many examples where some $S^{1}$ direction shrinks smoothly to zero,
starting with the bubble of nothing instability of the Kaluza-Klein vacuum
\cite{Witten:1981gj}. A similar solution with asymptotically locally AdS
boundary conditions was dubbed the AdS soliton \cite{Horowitz:1998ha}. These
solutions are easily constructed; they are simply double analytic
continuations of black hole solutions (in these examples, uncharged black
holes in asymptotically flat or asymptotically AdS space). In the AdS case,
the soliton is a time-independent solution, with a flat conformal boundary.
The spin structure of a spacetime with a contractible $S^{1}$ has antiperiodic
boundary conditions on the $S^{1}$. In the holographic AdS/CFT correspondence,
the dual theory thus has a spatial circle with antiperiodic boundary
conditions for fermions; if the CFT was supersymmetric this choice of boundary
conditions breaks the supersymmetry. The AdS soliton is dual to the ground
state in this theory. It has a negative energy, which is identified with the
Casimir energy of the dual CFT; the antiperiodic boundary conditions for
fermions spoil the cancellation between the bosons and fermions.

A simple generalization of these boundary conditions in the CFT is to add a
Wilson line around the non-trivial cycle, giving a non-trivial holonomy for
the bulk gauge field. A family of bulk solutions satisfying these boundary
conditions can be obtained by a double analytic continuation of electrically
charged black holes. After the analytic continuation, the solutions have a
magnetic flux through the circle; in the bulk the Wilson line is related to
the total amount of magnetic flux through this cycle. These are AdS analogues
of the Melvin fluxtube solutions in flat space \cite{Melvin:1963qx}, and were
previously studied in \cite{Astorino:2012zm,Lim:2018vbq,Kastor:2020wsm}. For
simplicity, we study the solutions primarily in four bulk dimensions with a
flat boundary, but we also consider generalisations in Euclidean signature to
a boundary which is $S^{1} \times\Sigma_{g}$ for some Riemann surface of genus
$g >1$, and to five bulk dimensions, which may be interesting for its relation
to $\mathcal{N}=4$ SYM.

Surprisingly, there is a solution of vanishing energy in this family which is
supersymmetric. This is surprising, as the solutions still have the same
structure as the AdS soliton, with a smooth origin which enforces antiperiodic
boundary conditions for the fermions on the shrinking $S^{1}$, which we
normally think of as breaking the supersymmetry. However, the holonomy around
the circle can compensate for the change in boundary conditions. The
supersymmetry of these solutions has been recognised in various guises before.
They are double analytic continuations of a set of supersymmetric naked
singularities identified in four dimensions in \cite{Romans:1991nq,
Caldarelli:1998hg} and in five dimensions in \cite{Gauntlett:2003fk}. They
were also studied in four dimensional Euclidean space very recently in
\cite{Bobev:2020pjk} as an extension of the supersymmetric black holes dual to
the twisted supersymmetric index. We add to these previous discussions by
emphasizing the previously unnoticed fact that these solutions are perfectly
well behaved Lorentzian configurations. Indeed, these configurations can be
thought as a supersymmetric extension of the AdS soliton of
\cite{Horowitz:1998ha}. We provide a detailed discussion of the role of its
spin structure; we will explicitly construct the antiperiodic Killing spinors
in the bulk spacetime.

In addition to these smooth soliton solutions, there are also solutions with
these boundary conditions where the $S^{1}$ does not have a smooth origin in
the bulk. In the simple case with flat boundary conditions, this solution is
simply AdS in Poincar\'e coordinates, with one of the spatial directions
identified. For $S^{1} \times\Sigma_{g}$ boundary conditions, the solution is
an analytic continuation of the magnetically charged black hole solution of
Romans, with an AdS$_{2} \times\Sigma_{g}$ near-horizon region
\cite{Romans:1991nq,Cacciatori:2009iz,Benini:2015eyy}. A solution with a
Wilson line is easily obtained by adding a constant gauge potential in the
bulk, without changing the geometry. These solutions are supersymmetric for
integer-quantized values of the Wilson line, with periodic Killing spinors for
even $n$ and antiperiodic Killing spinors for odd $n$.

Thus, for a range of values of the Wilson line on the $S^{1}$, there are two
bulk solutions with these boundary conditions: the magnetic flux tube solution
and the solution dressed with a constant gauge field.\footnote{As noted in
\cite{Kastor:2020wsm}, the magnetic flux solution exists only for a finite
range of values of the holonomy.} We will show that there is a phase
transition between these two families of solutions precisely at the
supersymmetric point with antiperiodic Killing spinors; the ground state of
the CFT with these boundary conditions is the soliton for smaller values of
the Wilson line, and the constant gauge field solution for larger values.

The path integral for the CFT on an $S^{1} \times\Sigma_{g}$ with the
supersymmetry-preserving choice of Wilson line computes the twisted
supersymmetric index \cite{Benini:2015noa,Benini:2016hjo,Closset:2016arn}%
.\footnote{See \cite{Zaffaroni:2019dhb} for a review and more references.} In
this context, both of these three-dimensional configurations have been
considered before: the solution with the constant gauge field was studied in
\cite{Benini:2015eyy} as the bulk dual of this twisted partition function. It
was noted in \cite{Bobev:2020pjk} that there is a supersymmetric solution with
an $\mathbb{R}^{2} \times\Sigma_{g}$ near-horizon region; our magnetic flux
tube. However, it was not previously appreciated that these solutions compete.

This is as far as we know the first example where there is a degeneracy
between bulk solutions in a supersymmetric partition function, and it would be
fascinating to understand its implications for the field theory. It would be
interesting to study the one-loop determinant of the bulk fields on both
solutions to see if this lifts the degeneracy. We conjecture that it does not,
and the supersymmetric ground state in field theory is dual to a superposition
of the two geometries. If so, this would be the first example we know where a
field theory ground state has such an interpretation. It would then be very
interesting to consider the observables in this state, such as correlation
functions, whose calculation in the bulk will be sensitive to the relative
phase between the two solutions. It is also interesting that there is
qualitative difference between even and odd integer-quantized Wilson lines; in
the even case, with periodic Killing spinors, there is a unique supersymmetric
solution in the bulk, while in the odd case there is this degeneracy.

In the first part of the paper, we will discuss the case with flat boundary in
AdS$_{4}$ in some detail. We set up the bulk theory in section \ref{theory},
and describe the solutions in section \ref{solitons}. We give a careful
analysis of supersymmetry in section \ref{susy}, and describe the phase
structure in section \ref{phase}. We then briefly discuss the extension to
$S^{1}\times\Sigma_{g}$ boundaries in section \ref{highg}. We generalize the
analysis to AdS$_{5}$ for flat boundaries in section
\ref{fived}.

\section{Four dimensions}

\subsection{Gauged $\mathcal{N}=2$ supergravity}

\label{theory}

Our explicit discussion of solutions in four dimensions is carried out in the
context of minimal gauged $\mathcal{N} = 2$ supergravity in four dimensions.
This theory was originally constructed in terms of the physical fields
\cite{Freedman:1976aw,Fradkin:1976xz}, whose supersymmetry transformations
only close under commutation up to equations of motion. Subsequently two
alternative constructions were presented based on the superconformal multiplet
calculus \cite{deWit:1980lyi,deWit:1984wbb,deWit:1984rvr}. This theory is a
universal subsector of a range of four-dimensional supersymmetric theories.
The bosonic sector is the familiar Einstein-Maxwell-AdS theory, with action
\begin{equation}
S\left(  g,A\right)  =\int d^{4}x\sqrt{-g}\left[  \frac{R}{2}-\frac{1}%
{8}F_{\mu\nu}F^{\mu\nu}+\frac{3}{\ell^{2}}\right]  , \label{Lag}%
\end{equation}
where $F(A)_{\mu\nu}=\partial_{\mu}A_{\nu}-\partial_{\nu}A_{\mu}\,.$The field
equations are%
\begin{align}
&  \partial_{\mu}\big(\sqrt{-g}F^{\mu\nu}\big)=0\,\,,\nonumber\\
&  R_{\mu\nu}-\tfrac{1}{2}g_{\mu\nu}R-\tfrac{1}{2}\big[F_{\mu\rho}\,F_{\nu}%
{}^{\rho}-\tfrac{1}{4}g_{\mu\nu}F_{\rho\sigma}F^{\rho\sigma}\big]-\frac
{3}{\ell^{2}}\,g_{\mu\nu}=0\,.
\end{align}
When supplemented by the fermionic sector, the supergravity theory is
invariant under supersymmetry transformations of all the fields. In our
calculations we will only make use of the transformation of the chiral
Rarita-Schwinger fields $\psi_{\mu}{\!}^{i}\,$:%
\begin{align}
\delta\psi_{\mu}{\!}^{i}=\;  &  2\,\mathcal{D}_{\mu}\epsilon^{i}-\tfrac{1}%
{4}F(A)_{\rho\sigma}\gamma^{\rho\sigma}\gamma_{\mu}\,\varepsilon
^{ij}\,\epsilon_{j}+\ell^{-1}\,\varepsilon^{ij}\,t_{j}{}^{k}\,\gamma_{\mu
}\epsilon_{k}\,,\\[1mm]
\delta\psi_{\mu\,i}=\;  &  2\,\mathcal{D}_{\mu}\epsilon_{i}-\tfrac{1}%
{4}F(A)_{\rho\sigma}\gamma^{\rho\sigma}\gamma_{\mu}\,\varepsilon
_{ij}\,\epsilon^{j}+\ell^{-1}\,\varepsilon_{ij}\,t^{j}{}_{k}\,\gamma_{\mu
}\epsilon^{k}\,,
\end{align}
where $\gamma^{5}=-\mathrm{i}\gamma^{0}\gamma^{1}\gamma^{2}\gamma^{3}$,
$\gamma^{5}\psi_{\mu}^{i}=\psi_{\mu}^{i}$ and $\gamma^{5}\psi_{\mu i}%
=-\psi_{\mu i}$ and $i=1,2$ and we shall pick $t^{i}{}_{j}=$ $\mathrm{i}%
\sigma_{3}\Longrightarrow t_{i}{}^{j}=$ $-\mathrm{i}\sigma_{3}$. The covariant
derivatives of the supersymmetry parameters are given by
\begin{align}
\mathcal{D}_{\mu}\epsilon^{i}=  &  \;\big(\partial_{\mu}+\tfrac{1}{4}%
\omega_{\mu}{\!}^{ab}\gamma_{ab}\big)\epsilon^{i}-\tfrac{1}{2\ell}A_{\mu
}\,t^{i}{\!}_{j}\,\epsilon^{j}\,,\label{eq:covariant-epsilon-der}\\[1mm]
\mathcal{D}_{\mu}\epsilon_{i}=  &  \;\big(\partial_{\mu}+\tfrac{1}{4}%
\omega_{\mu}{\!}^{ab}\gamma_{ab}\big)\epsilon_{i}-\tfrac{1}{2\ell}A_{\mu
}\,t_{i}{}^{j}\,\epsilon_{j}\,.
\end{align}
In our calculations we find that is convenient to work with the spinor%
\begin{equation}
\chi\equiv\epsilon^{1}+\epsilon_{2}\,, \label{eq:complex-susy}%
\end{equation}
such that the Killing spinor equation is
\begin{equation}
2\,\mathcal{D}_{\mu}\chi+\tfrac{1}{4}F(A)_{\rho\sigma}\gamma^{\rho\sigma
}\gamma_{\mu}\gamma^{5}\,\chi-\ell^{-1}\mathrm{i}\,\gamma_{\mu}\gamma^{5}%
\chi=0\,, \label{eq:complex-killing-spinor-eqs}%
\end{equation}
where the covariant derivative of $\chi$ follows from
\eqref{eq:covariant-epsilon-der},
\begin{equation}
\mathcal{D}_{\mu}\chi=\big(\partial_{\mu}+\tfrac{1}{4}\omega_{\mu}{\!}%
^{ab}\gamma_{ab}-\tfrac{1}{2\ell}\,\mathrm{i}\,A_{\mu}\big)\chi\,.
\label{eq:cov-der-chi}%
\end{equation}
We recall that all fermionic fields have been suppressed on the right-hand
side of equation \eqref{eq:cov-der-chi}, because we will be dealing with
purely bosonic backgrounds. In the following sections we will consider a class
of soliton solutions that can be partially supersymmetric. Their possible
supersymmetry will be investigated by analyzing the equation \eqref{eq:complex-killing-spinor-eqs}.

The Lagrangian \eqref{Lag} can be obtained from the compactification of eleven
dimensional supergravity over the seven sphere with the ansatz
\cite{Cvetic:1999xp}
\begin{align}
ds_{11}^{2}  &  =ds_{4}^{2}+4\ell^{2}\sum_{i}\left(  d\mu_{i}^{2}+\mu_{i}%
^{2}\left(  d\phi_{i}+\frac{1}{4\ell}A\right)  ^{2}\right)  .\label{11d}\\
F_{4}  &  =-\frac{3}{\ell}\epsilon_{4}-2\ell^{2}\sum_{i}\mu_{i}d\mu_{i}%
\wedge\left(  d\phi_{i}+\frac{1}{4\ell}A\right)  \wedge\ast_{4}dA
\end{align}
where $\ast_{4}$ is the Hodge dual with respect to the four-dimensional metric
$ds_{4}^{2}$ and $\epsilon_{4}$ its volume form. The $\phi_{i}$ are $2\pi$
periodic angular coordinates parametrizing the four independent rotations on
$S^{7}$. We will be interested in considering the higher-dimensional
interpretation of some of our solutions using this uplift.

\subsection{Planar solitons}

\label{solitons}

The solutions we consider all have metric
\begin{equation}
ds^{2}=\frac{r^{2}}{\ell^{2}}\left(  -dt^{2}+dz^{2}\right)  +\frac{dr^{2}%
}{f(r)}+f(r)d\phi^{2}\,. \label{bhm}%
\end{equation}
When considering supersymmetry we will also need a corresponding set of
vierbeine, which we take to be
\begin{align}
e^{0}  &  =\frac{r}{\ell}dt\,,\nonumber\\[0.5mm]
e^{1}  &  =\frac{\,dr}{\sqrt{f(r)}}\,,\nonumber\\[0.3mm]
e^{2}  &  =\sqrt{f(r)}d\phi\,,\nonumber\\[2.1mm]
e^{3}  &  =\frac{r}{\ell}dz.
\end{align}

The simplest solution has
\begin{equation}
f(r)=\frac{r^{2}}{\ell^{2}}\, .
\end{equation}
If $\phi$ is not periodically identified, this is simply pure $\mathrm{AdS}%
_{4}$ in Poincar\'e coordinates. However, we are interested in considering
solutions with $\phi$ periodically identified in the boundary. For
Poincar\'e-AdS, we can impose this identification on the bulk as a quotient.
We will postpone a full discussion of this quotient until we have discussed
the soliton solution.

The soliton solution is
\begin{align}
f(r)  &  =\frac{r^{2}}{\ell^{2}}-\frac{\mu}{r}-\frac{Q^{2}}{r^{2}}\,,\\
A  &  =\left(  \frac{2Q}{r}-\frac{2Q}{r_{0}}\right)  d\phi\text{ ,}%
\end{align}
where $r_{0}$ is the largest root of the equation $f(r_{0})=0$.\footnote{Note
that $f$ always has at least one positive root, for all real values of $\mu
,Q$.} This solution can be obtained by a double analytic continuation from an
electrically charged black hole solution, where the analytic continuation also
involves analytically continuing the charge $Q$. We have $r \in\left[
r_{0},\infty\right)  $, and regularity of the metric at $r=r_{0}$ requires
that $\phi$ is periodic with period
\begin{equation}
\Delta\phi=\frac{4\pi\ell^{2}r_{0}^{3}}{3r_{0}^{4}+Q^{2}\ell^{2}} . \label{HP}%
\end{equation}
We have added a constant contribution to the gauge potential to ensure
regularity at $r=r_{0}$; there is then a non-trivial holonomy of the gauge
field at infinity. This is related to the net magnetic flux along the $z$
axis,
\begin{equation}
\Phi= - \oint A_{\phi}\left(  r=\infty\right)  d\phi= - \frac{1}{2}\int
F_{\mu\nu}dx^{\mu}\wedge dx^{\nu} =\frac{2Q}{r_{0}} \Delta\phi.
\end{equation}

The soliton solution is determined by two parameters; from the bulk
perspective the natural parametrization is $\mu,Q$ or $r_{0},Q$. From the
boundary perspective, the natural parameters are $\Delta\phi, \Phi$. In the
usual holographic dictionary, we fix the boundary geometry and the boundary
value of $A_{\mu}$, which is interpreted as a background gauge field coupled
to a global $U(1)$ symmetry of the CFT. Recall that as discussed in the
introduction, for the soliton solutions the bulk spin structure is
antiperiodic on the $\phi$ circle, as this contracts smoothly in the interior.

Thus, we consider the CFT on a background
\begin{equation}
ds_{Boundary}^{2}=-dt^{2}+dz^{2}+d\phi^{2},
\end{equation}
with $\phi$ taken to be periodic with period $\Delta\phi$, and an antiperiodic
spin structure for the fermions, and a Wilson line (holonomy) on the $\phi$
circle with value $\Phi$, and we look for bulk solutions satisfying these
boundary conditions. We obtain a suitable bulk soliton solution by solving for
$\mu, Q$ as functions of $\Delta\phi, \Phi$. We have
\begin{equation}
\mu= \frac{r_{0}^{4} - Q^{2} \ell^{2}}{r_{0} \ell^{2}}, \quad Q = \frac{ \Phi
r_{0}}{2 \Delta\phi},
\end{equation}
and inverting \eqref{HP} gives \cite{Kastor:2020wsm}
\begin{equation}
r_{0} = \frac{2\pi\ell^{2}}{3 \Delta\phi} \left(  1 \pm\sqrt{ 1 - \frac
{\Phi^{2}}{\Phi_{max}^{2}}} \right)  ,
\end{equation}
where $\Phi_{max} = \frac{4\pi}{\sqrt{3}} \ell$. We see that there are two
solutions for $r_{0}$ for $\Phi\in[0, \Phi_{max}]$. At $\Phi=0$, the minus
branch has $r_{0} =0$, giving $\mu=Q=0$, so it reduces to the Poincar\'e-AdS
solution. The plus branch has $Q=0$, and
\begin{equation}
\mu= \left(  \frac{4 \pi\ell}{3 \Delta\phi} \right)  ^{3} \ell,
\end{equation}
which is just the AdS soliton. The two branches coalesce at $\Phi= \Phi_{max}$.

The parameters $\mu, Q$ control the subleading parts of the metric and gauge
field asymptotically, so they can be interpreted as vevs of the corresponding
operators in the field theory. The gauge field gives a vev for the current in
the boundary theory,
\begin{equation}
\langle J_{\phi}\rangle= 2Q,
\end{equation}
and the metric gives a vev for the stress tensor,
\begin{equation}
\left\langle T_{tt}\right\rangle =-\frac{\mu}{2\ell^{2}}\text{ }%
,\qquad\left\langle T_{zz}\right\rangle =\frac{\mu}{2\ell^{2}}\text{ }%
,\qquad\left\langle T_{\phi\phi}\right\rangle =-\frac{\mu}{\ell^{2}}\text{ }.
\end{equation}
We see that the energy density of the soliton solutions is negative when $\mu$
is positive. We can write explicitly \cite{Kastor:2020wsm}
\begin{equation}
\left\langle T_{tt}\right\rangle = - \left(  \frac{4\pi\ell}{3 \Delta\phi}
\right)  ^{3} \frac{1}{4 \ell} \left[  1 - \frac{3 \Phi^{2}}{2 \Phi_{max}^{2}}
\pm\left(  1 - \frac{\Phi^{2}}{\Phi_{max}^{2}} \right)  ^{3/2} \right]  .
\end{equation}
We see that the minus branch always has $\left\langle T_{tt}\right\rangle
\geq0$, and the plus branch has $\left\langle T_{tt}\right\rangle < 0$ for
$\Phi\in[0,\Phi_{S})$, where
\begin{equation}
\Phi_{S} = \frac{\sqrt{3}}{2} \Phi_{max} = 2\pi\ell.
\end{equation}

In addition to the soliton solutions, another solution with these boundary
conditions is simply to take the Poincar\'e-AdS solution, with a constant
gauge field $A = - \frac{2 Q}{r_{0}} d\phi$. If $\phi$ were not periodic, this
is just pure AdS in Poincar\'e coordinates, and the gauge field is pure gauge.
Taking $\phi$ to be periodic with period $\Delta\phi$, the gauge field has a
constant Wilson loop $\Phi= - \frac{2Q}{r_{0}} \Delta\phi$ for all $r$, which
can't be set to zero by a gauge transformation.

It is useful to think of this solution as a quotient of AdS$_{4} \times S^{7}%
$. For $\Phi=0$, the quotient has fixed points at $r=0$, so this is not a
smooth solution. Furthermore, for the antiperiodic boundary conditions we
consider, this solution is unstable as a solution of string theory, as a
string wrapped around the $\phi$ circle will become tachyonic for sufficiently
small $r$. This solution decays by tachyon condensation, whose likely endpoint
is the AdS soliton \cite{Adams:2001sv,Horowitz:2006mr}.

However, for $\Phi\neq0$, the identification involves a shift of the angular
coordinates on the sphere. The solution with a constant gauge field uplifts as
in \eqref{11d}. Since the gauge potential is constant, if $\phi$ is not
periodically identified we can eliminate the gauge field by coordinate
redefinitions $\tilde\phi_{i} = \phi_{i} + \frac{A_{\phi}}{4\ell} \phi$. The
periodic identification of $\phi$ at fixed $\phi_{i}$ then acts as $(\phi,
\tilde\phi_{i}) \sim(\phi+ \Delta\phi, \tilde\phi_{i} + \Phi/4\ell)$. This is
now a smooth quotient, with no fixed points,\footnote{More carefully, the
identification is free of fixed points so long as $\Phi\neq8\pi\ell n$ for
integer $n$, so that the action on the $\phi_{i}$ coordinates is globally
non-trivial.} and the circle in the higher-dimensional space has a minimum
size $\Phi/8$, so for $\Phi$ larger than the string scale the tachyon is
lifted, and this is a physical solution. It is then interesting to compare
this solution to the soliton. We will discuss the phase structure in section
\ref{phase}, but we first discuss in more detail the supersymmetric solutions
at the special value $\Phi= \Phi_{S}$.

\subsection{Supersymmetric solutions}

\label{susy}

The configuration with zero energy, $\mu=0$, which occurs on the plus branch
of solitons at $\Phi=\Phi_{S}=2\pi\ell$, is supersymmetric. At this value of
$\Phi$, the Poincar\'{e}-AdS solution is also supersymmetric. The
supersymmetry of the soliton in the Euclidean section has been noted before
\cite{Bobev:2020pjk}, but we give a slightly more careful analysis taking into
account the antiperiodic boundary conditions for the fermions on the $\phi$
circle. Our remark is that the supersymmetric soliton solutions are perfectly
well behaved in the Lorentzian section. We will explicitly exhibit the Killing
spinors for this solution, and see how they arise from the higher-dimensional perspective.

For the solitons with $\mu=0$, we can write $r_{0}\bigskip=\sqrt
{\ell\left\vert Q\right\vert }$. The form of the solution can be simplified by
a coordinate transformation
\begin{equation}
r=r_{0}\bigskip\sqrt{\cosh(\rho)},
\end{equation}
where the configuration reads%
\begin{equation}
ds^{2}=\frac{\ell^{2}}{4}d\rho^{2}+\frac{|Q|}{\ell}\left[  \cosh\left(
\rho\right)  \left(  -dt^{2}+dz^{2}\right)  +\frac{\sinh\left(  \rho\right)
^{2}}{\cosh(\rho)}d\phi^{2}\right]  ,
\end{equation}%
\begin{equation}
A=\frac{2Q}{r_{0} }\left(  \frac{1}{\sqrt{\cosh\rho}}-1\right)  d\phi.
\end{equation}
We use a Majorana basis for the Clifford algebra%
\begin{equation}
\gamma^{0}=-\mathrm{i}\left(
\begin{array}
[c]{cc}%
0 & \sigma_{2}\\
\sigma_{2} & 0
\end{array}
\right)  , \quad\gamma^{1}=-\left(
\begin{array}
[c]{cc}%
\sigma_{3} & 0\\
0 & \sigma_{3}%
\end{array}
\right)  , \quad\gamma^{2}=\mathrm{i}\left(
\begin{array}
[c]{cc}%
0 & -\sigma_{2}\\
\sigma_{2} & 0
\end{array}
\right)  , \quad\gamma^{3}=\left(
\begin{array}
[c]{cc}%
\sigma_{1} & 0\\
0 & \sigma_{1}%
\end{array}
\right)  .
\end{equation}

There are two independent solutions to the Killing spinor equation on this
background:%
\begin{equation}
\chi_{1}=\frac{\exp\left(  -\mathrm{i}\pi\frac{\phi}{\Delta\phi}\right)
}{\cosh\left(  \rho\right)  ^{1/4}}\left(
\begin{array}
[c]{c}%
\sinh\frac{\rho}{2}\\
-\cosh\frac{\rho}{2}\\
\mathrm{i}\cosh\frac{\rho}{2}\\
\mathrm{i}\sinh\frac{\rho}{2}%
\end{array}
\right)  ,
\end{equation}%
\begin{equation}
\chi_{2}=\frac{\exp\left(  -\mathrm{i}\pi\frac{\phi}{\Delta\phi}\right)
}{\cosh\left(  \rho\right)  ^{1/4}}\left(
\begin{array}
[c]{c}%
-\cosh\frac{\rho}{2}\\
\sinh\frac{\rho}{2}\\
\mathrm{i}\sinh\frac{\rho}{2}\\
\mathrm{i}\cosh\frac{\rho}{2}%
\end{array}
\right)  .
\end{equation}
The two spinors $\chi_{1}$ and $\chi_{2}$, can be projected over four chiral
spinors and therefore the soliton solution is 1/2 BPS. They close on the
following Killing vectors:%
\begin{equation}
\chi_{2}^{\dagger}\gamma^{0}\gamma^{\mu}\chi_{2}\partial_{\mu}=\chi
_{1}^{\dagger}\gamma^{0}\gamma^{\mu}\chi_{1}\partial_{\mu}=\frac{2\ell}{r_{0}%
}\partial_{t},
\end{equation}%
\[
\chi_{1}^{\dagger}\gamma^{0}\gamma^{\mu}\chi_{2}\partial_{\mu}=-\frac{2\ell
}{r_{0}}\left(  \partial_{\phi}+\mathrm{i}\partial_{z}\right)  \text{ }.
\]
There is still the issue that the spinors $\chi_{1}$ and $\chi_{2}$ seem to be
ill-defined at $\rho=0$, as they have an explicit dependence on $\phi$,
however this dependence corresponds exactly to the form of the Killing spinors
of Minkowski in cylindrical coordinates:%

\begin{equation}
\left.  \chi_{1}\right\vert _{\rho=0}=\exp\left(  -\mathrm{i}\pi\frac{\phi
}{\Delta\phi}\right)  \left(
\begin{array}
[c]{c}%
0\\
-1\\
\mathrm{i}\\
0
\end{array}
\right)  ,
\end{equation}%
\begin{equation}
\chi_{2}=\exp\left(  -\mathrm{i}\pi\frac{\phi}{\Delta\phi}\right)  \left(
\begin{array}
[c]{c}%
-1\\
0\\
0\\
\mathrm{i}%
\end{array}
\right)  .
\end{equation}

Hence, our Killing spinors interpolate smoothly between half of the Killing
spinors of Minkowski and half of the Killing spinors of AdS$_{4}$. Indeed, we
see that we have solutions of the Killing spinor equation with non-trivial
dependence on $\phi$, whereas we would normally expect the Killing spinors to
be constant along these flat directions. This satisfies the Killing spinor
equation \eqref{eq:complex-killing-spinor-eqs} due to a cancellation between
the derivative $\partial_{\phi}$ and the gauge potential $A_{\phi}$ in the
covariant derivative $\mathcal{D}_{\phi}$.

For $\Phi= \Phi_{S} = 2\pi\ell$, the Poincar\'e -AdS solution is also
supersymmetric. This follows trivially from the fact that it gives the large
$r$ asymptotics of the supersymmetric soliton solution; if the soliton
preserves some supersymmetry, the metric it approaches at large $r$ must
preserve at least the same amount of supersymmetry. Taking the large $\rho$
limit gives
\begin{equation}
\chi_{1} =\exp\left(  -\mathrm{i}\pi\frac{\phi}{\Delta\phi}\right)
e^{\frac{\rho}{4}} \left(
\begin{array}
[c]{c}%
1\\
-1\\
\mathrm{i}\\
\mathrm{i}%
\end{array}
\right)  ,
\end{equation}
\begin{equation}
\chi_{2} =\exp\left(  -\mathrm{i}\pi\frac{\phi}{\Delta\phi}\right)
e^{\frac{\rho}{4}} \left(
\begin{array}
[c]{c}%
-1\\
1\\
\mathrm{i}\\
\mathrm{i}%
\end{array}
\right)  .
\end{equation}

On pure AdS$_{4}$, the theory has four Killing spinors; if we introduce a
Wilson loop on this background with $A=A_{\phi}d\phi$, we have that the local
solutions to the Killing spinor equation are%
\begin{equation}
\chi_{1}^{AdS}=\exp\left(  \frac{\mathrm{i}A_{\phi}}{2\ell}\phi\right)
r^{1/2}\left(
\begin{array}
[c]{c}%
1\\
-1\\
0\\
0
\end{array}
\right)  \text{,}\qquad\chi_{2}^{AdS}=\exp\left(  \frac{\mathrm{i}A_{\phi}%
}{2\ell}\phi\right)  r^{1/2}\left(
\begin{array}
[c]{c}%
0\\
0\\
1\\
1
\end{array}
\right)  \text{ ,}%
\end{equation}%
\begin{equation}
\chi_{3}^{AdS}=\exp\left(  \frac{\mathrm{i}A_{\phi}}{2\ell}\phi\right)
\left(
\begin{array}
[c]{c}%
r^{1/2}\left(  t+\phi\right) \\
-r^{1/2}\left(  t+\phi\right) \\
-r^{1/2}z-\ell^{2}r^{-1/2}\\
-r^{1/2}z+\ell^{2}r^{-1/2}%
\end{array}
\right)  \text{,}\qquad\chi_{4}^{AdS}=\exp\left(  \frac{\mathrm{i}A_{\phi}%
}{2\ell}\phi\right)  \left(
\begin{array}
[c]{c}%
-r^{1/2}z+\ell^{2}r^{-1/2}\\
r^{1/2}z+\ell^{2}r^{-1/2}\\
r^{1/2}\left(  t-\phi\right) \\
r^{1/2}\left(  t-\phi\right)
\end{array}
\right)  \,.
\end{equation}
The second two spinors, $\chi_{3}^{AdS}$ and $\chi_{4}^{AdS}$, are not
invariant under the identification which makes $\phi$ periodic, so this breaks
at least half the supersymmetry. The first two, $\chi_{1}^{AdS}$ and $\chi
_{2}^{AdS}$, are invariant (up to sign) if $A_{\phi}=2\pi n\ell/\Delta\phi$,
so $\Phi=2\pi n\ell$, for integer $n$. Thus, the identified Poincar\'{e}-AdS
solution with $\phi$ periodic and $\Phi=2\pi\ell$ has the same supersymmetry
as the soliton solution. The preserved supersymmetries correspond to the
Poincar\'{e} supersymmetries of the boundary theory, while the broken ones
correspond to superconformal symmetries of the boundary theory, which are
broken as the periodic identification introduces a choice of scale, breaking
the conformal symmetry.

It is also instructive to understand how the dependence on $\phi$ arises from
a higher-dimensional perspective. Thought of as solutions on AdS$_{4}\times
S^{7}$, the Killing spinors corresponding to Poincar\'{e} supersymmetries have
the structure
\begin{equation}
\chi\sim F(r)G(\mu_{i})e^{-\frac{\mathrm{i}}{2}(\tilde{\phi}_{1}+\tilde{\phi
}_{2}+\tilde{\phi}_{3}+\tilde{\phi}_{4})}\chi_{0}, \label{11ks}%
\end{equation}
where $\chi_{0}$ is a constant spinor. The dependence on the $\tilde{\phi}%
_{i}$ is fixed by the fact that these are contractible cycles on $S^{7}$, so
all fermions are antiperiodic around them. Thus, under the quotient action
$(\phi,\tilde{\phi}_{i})\sim(\phi+\Delta\phi,\tilde{\phi}_{i}+\Phi/4\ell)$
with $\Phi=\Phi_{S}=2\pi\ell$, these spinors are antiperiodic, $\chi
\rightarrow-\chi$, and they survive the identification, as we are keeping
precisely the sector of spinors which are antiperiodic under the quotient
action.\footnote{Note we have considered a quotient action which acts in the
same way on all the $\phi_{i}$ because we are focusing for simplicity on the
minimal gauged supergravity in four dimensions, but this supersymmetry
argument requires only that the sum of the shifts $\sum_{i}\Delta\tilde{\phi
}_{i}=2\pi$.} Another way to say this is to rewrite $\tilde{\phi}_{i}=\phi
_{i}+\frac{\pi}{2}\frac{\phi}{\Delta\phi}$, which gives
\begin{equation}
\chi\sim F(r)G(\mu_{i})e^{-\mathrm{i}\pi\frac{\phi}{\Delta\phi}}%
e^{-\frac{\mathrm{i}}{2}(\phi_{1}+\phi_{2}+\phi_{3}+\phi_{4})}\chi_{0}.
\end{equation}
Dimensionally reducing over the $\phi_{i}$ then gives the Killing spinors with
a Wilson line written above.

Thus, for $\Phi=\Phi_{S}$, we have two supersymmetric solutions; the soliton
with $\mu=0$ and the quotient of AdS. We next turn to the comparison between
these two solutions.

\subsection{Phase diagram}

\label{phase}

We now consider the phase transitions between the different solutions. We
consider the standard AdS boundary conditions fixing the leading asymptotic
falloff of the metric and gauge field $A$. The boundary conditions are then
parametrized by $\Delta\phi$ and $\Phi$. We note that the $U(1)$ gauge field,
and hence $\Phi$, take values in a circle. This is most evident from the
perspective of the uplift \eqref{11d}; a shift of $\Phi\to\Phi+ 8 \pi\ell n$
for integer $n$ shifts the $\phi_{i}$ by $2\pi n$. The range of physically
inequivalent values is thus $\Phi\in[-4\pi\ell, 4\pi\ell)$. We can restrict
attention to positive $\Phi$, as we have done so far; corresponding results
for negative $\Phi$ are obtained by reversing the sign of $Q$ for the soliton.
For $\Phi\in[0, \Phi_{max}]$, there are three different solutions for each
choice of $\Delta\phi$ and $\Phi$; the two branches of solitons and
Poincar\'e-AdS with the appropriate identification on $\phi$. For $\Phi
\in(\Phi_{max}, 4\pi\ell]$, we just have Poincar\'e-AdS.

\begin{figure}[th]
\centering
\includegraphics[scale=0.3]{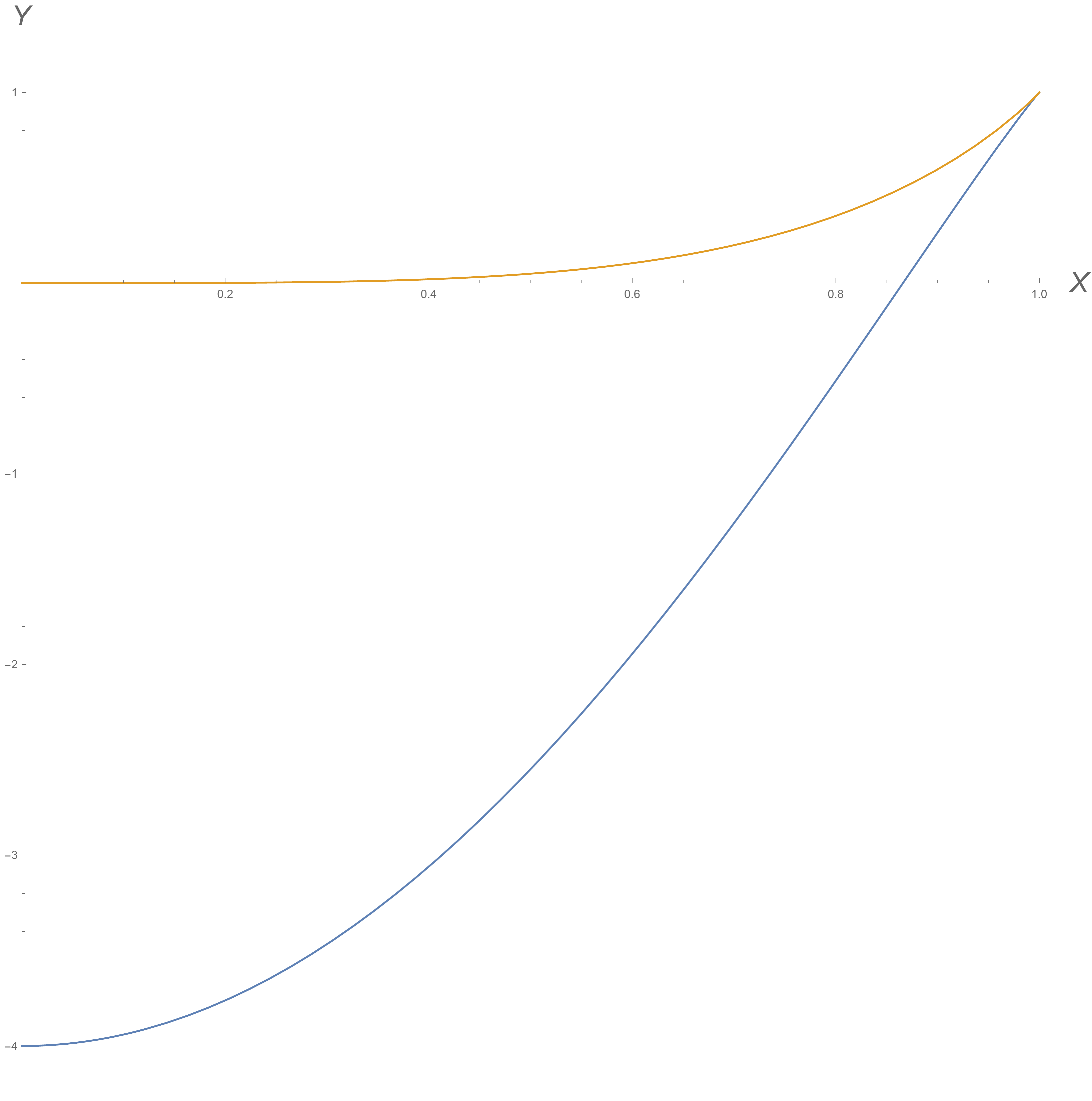}\caption{The rescaled energy density,
$Y= \frac{27 \Delta\phi^{3}}{8\pi^{3}\ell^{2}}\langle T_{tt} \rangle$, vs the
rescaled magnetic flux, $X=\Phi/\Phi_{max}$, for the two branches of soliton
solutions. The lower branch crosses the axis at $X_{S} = \Phi_{S}/\Phi_{max} =
\sqrt{3}/2$.}%
\label{figure1}%
\end{figure}

In Lorentzian signature, these solutions should all be dual to some states in
the dual CFT. For $\Phi=0$, the solution on the lower soliton branch is the
AdS soliton, and this is conjectured to be dual to the ground state of the CFT
\cite{Horowitz:1998ha}. It is reasonable to assume that the lowest-energy
solution remains dual to the ground state as we turn on $\Phi$. The energy for
the soliton solutions is plotted in figure \ref{figure1}. For $\Phi\in[0,
\Phi_{S})$, the lowest energy solution is the lower soliton branch. At $\Phi=
\Phi_{S}$, the two supersymmetric solutions discussed in the previous section
both have zero energy, so they are degenerate. For $\Phi\in(\Phi_{S}, 4
\pi\ell]$, Poincar\'e-AdS is the lowest-energy solution.

\bigskip

As for the AdS soliton, the energy $\langle T_{tt} \rangle= - \frac{\mu}%
{2\ell^{2}}$ should be interpreted as a Casimir energy of the ground state. We
see that turning on a Wilson line initially increases (reduces the magnitude
of) the Casimir energy, presumably by deforming the spectrum of fields charged
under this $U(1)_{R}$ symmetry. It is surprising that for $\Phi> \Phi_{S}$ the
energy is zero, independent of the Wilson line $\Phi$. The cancellation
between bosons and fermions is restored at the supersymmetric point, but it's
unexpected that this continues for higher values of the Wilson line. It would
be interesting to investigate this directly from the field theory perspective.

The other mystery is the precise nature of the ground state at $\Phi= \Phi
_{S}$. In our supergravity analysis, we have two degenerate solutions. This is
the first example we know of with distinct supersymmetric solutions with the
same boundary conditions, which are degenerate in energy. It is possible that
this degeneracy is lifted by subleading effects, for example by considering
the Casimir energy for bulk fields on these two geometries. This is certainly
an interesting issue to investigate, but it seems unlikely that this will lift
the degeneracy: the bulk solutions are supersymmetric, so the natural guess is
that contributions to the bulk Casimir from bosonic and fermionic fields will
cancel. Also, there is a phase transition between the two solutions which is
at least very near the supersymmetric point; it would be surprising if it were
then not precisely at this point. The alternative possibility is that the
supersymmetric ground state in the field theory is dual to a superposition of
these two geometries. This would be very interesting if true: the calculation
of observables, such as correlation functions, in this ground state, would
then involve contributions from both geometries. If these observables could be
calculated on the CFT side, they would give a unique probe of relative phase
information in the superposition of bulk geometries.

It is also interesting to note that the Poincar\'e-AdS solution is
supersymmetric for $\Phi= 2\pi n \ell$. Since $\Phi$ is periodic with period
$8 \pi\ell$, there are four physically inequivalent values giving
supersymmetric solutions, which we can take to be $n=-1,0,1,2$. For odd $n$,
the unbroken Killing spinors in Poincar\'e-AdS are antiperiodic on the $\phi$
circle, while for even $n$, they are periodic. For the odd cases, there is a
degenerate soliton solution, but for the even cases Poincar\'e-AdS is the only
supersymmetric solution. It would be interesting to understand this difference
between odd and even $n$ from the field theory perspective.

\subsection{Euclidean partition function}

It is also interesting to consider the phase structure from a Euclidean
perspective, to make contact with previous work on supersymmetric partition
functions \cite{Benini:2015noa,Benini:2016hjo,Zaffaroni:2019dhb}. Previous
work on this area has focused on the Euclidean theory with an $S^{1}
\times\Sigma_{g}$ boundary, where $\Sigma_{g}$ is a Riemann surface of genus
$g >1$, but this was extended to consider genus 1 in \cite{Bobev:2020pjk},
corresponding to the case we have considered so far. We will comment on the
extension of our discussion to higher genus in section \ref{highg}.

The Euclidean continuation of our solutions are solutions with an $S^{1}
\times\mathbb{R}^{2}$ boundary. These are usually treated thinking of the
$S^{1}$ as the Euclidean time circle, so $\Phi$ is a chemical potential for
the R-charge of the theory and the boundary path integral is interpreted as
the partition function $Z = \mathrm{Tr}\, e^{i \Phi J_{R}} e^{-\Delta\phi\,
H}$ of the CFT on $\mathbb{R}^{2}$. For $\Phi= 2\pi\ell$, this partition
function becomes a supersymmetric index.

In the saddle-point approximation, $Z \approx e^{-S_{E}}$, where $S_{E}$ is
the action of the Euclidean solutions. For the solitons, the Wick rotated
metric is
\begin{equation}
ds^{2}=\frac{r^{2}}{\ell^{2}}\left(  d\tau^{2}+dz^{2}\right)  +\frac{dr^{2}%
}{f(r)}+f(r)d\phi^{2}\,,
\end{equation}
with $f(r)=\frac{r^{2}}{\ell^{2}}-\frac{Q^{2}}{r^{2}}-\frac{\mu}{r}$ as
before. The Euclidean action is
\begin{align}
\frac{S_{E}}{V}  &  =-\lim_{R\rightarrow\infty}\left[  \left(  \int_{r_{0}%
}^{R}dr\sqrt{g_{E}}\frac{R}{2}-\frac{1}{8}F_{\mu\nu}F^{\mu\nu}+\frac{3}%
{\ell^{2}}\right)  +\left(  K\sqrt{h}-\frac{2}{\ell}\sqrt{h}\right)
_{r=R}\right] \nonumber\\
&  =-\frac{\mu}{2\ell^{2}},
\end{align}
where $V$ is the coordinate volume element on the boundary, $V=\int d\tau
d\phi dz$. The Poincar\'e-AdS solution with a constant Wilson line always has
zero action. Since the Euclidean action is proportional to $\mu$, the phase
structure is the same as in our Lorentzian discussion above; for $\Phi\in[0,
\Phi_{S})$, the dominant saddle-point is the lower soliton branch, for
$\Phi\in(\Phi_{S}, 4 \pi\ell]$ it is Poincar\'e-AdS. There is a phase
transition at $\Phi=\Phi_{S}$.

For the supersymmetric case $\Phi=\Phi_{S}$, the partition function is an
index, which can be computed by supersymmetric localization. For the present
case, the CFT calculation gives an answer which vanishes to leading order in
$N$ \cite{Benini:2015eyy}, in agreement with the bulk calculation. This
calculation has previously been matched to the bulk result from
Poincar\'e-AdS; we observe that both bulk saddles match the CFT result, and as
discussed previously it will be interesting to understand further what the
precise bulk dual of the index is.

An interesting issue here is the calculation of the expectation value of the
$R$ charge in the ensemble. In the CFT, we can calculate the partition
function exactly only at $\Phi=\Phi_{S}$, so we can't extract a value for the
R-charge from this; thermodynamically, to calculate the R-charge we need to
consider the variation $\frac{\partial Z}{\partial\Phi}$. Holographically, we
can evaluate the partition function for general $\Phi$, but at the
supersymmetric point the degeneracy prevents us from evaluating the R-charge.
The soliton saddle has $\langle J_{\phi}\rangle= 2Q$, corresponding to a
non-zero R-charge density, while Poincar\'e-AdS has $\langle J_{\phi}%
\rangle=0$. Since they are degenerate at $\Phi=\Phi_{S}$, we don't know what
value to use. Thermodynamically, the phase transition implies $Z$ is not a
smooth function of $\Phi$, so $\frac{\partial Z}{\partial\Phi}$ is ill-defined
at $\Phi=\Phi_{S}$.

\subsection{Fixed charge}

Since we are in four bulk dimensions, we can also define a holographic
correspondence with alternative boundary conditions, where we fix $F^{\nu
}\equiv\sqrt{h} n_{\mu}F^{\mu\nu}$ at the boundary (where $n_{\mu}$ is the
unit normal to the boundary), instead of fixing $A_{\mu}$ \cite{Marolf:2006nd}%
. This corresponds to fixing $\Delta\phi$ and $Q$. With these boundary
conditions, Poincar\'e-AdS is not a solution. For the soliton solutions, we
can solve \eqref{HP} to determine $r_{0}$ as a function of $\Delta\phi$ and
$Q$. We can see that $\Delta\phi$ is small at both $r_{0} \to0$ and $r_{0}
\to\infty$, with a maximum at $r_{0}^{4} = Q^{2} \ell^{2}$, which implies
$\mu=0$, where $\Delta\phi= \Delta\phi_{max} = \pi\sqrt{ \frac{\ell^{3}}{Q}}$.
Thus, the supersymmetric solution in this case corresponds to the maximum
value of $\Delta\phi$.

\begin{figure}[th]
\centering
\includegraphics[scale=0.3]{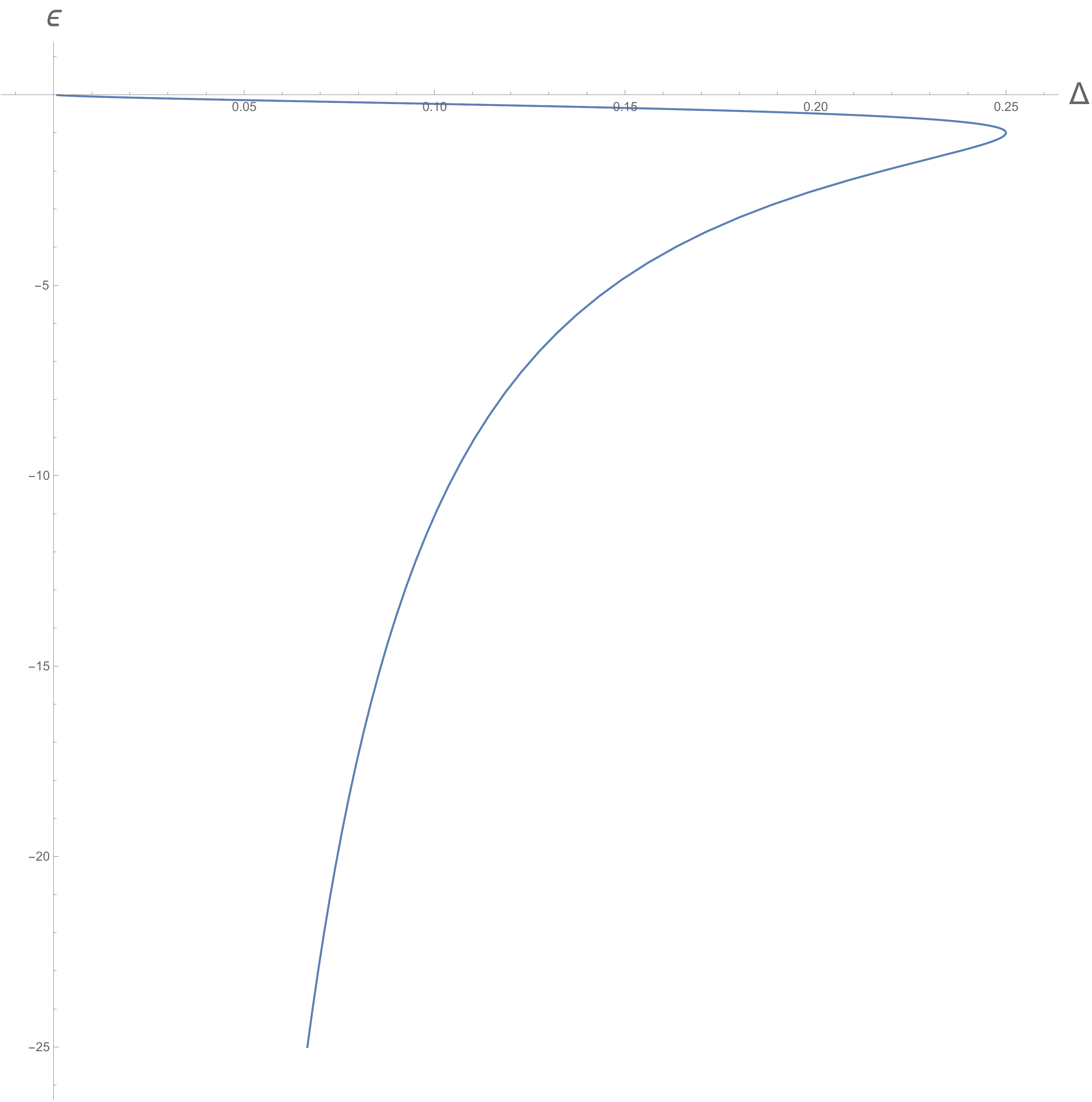}\caption{The action $\epsilon
=\frac{\ell}{2\pi\left\vert Q\right\vert }S^{\prime}/V$ vs $\Delta=\frac
{\sqrt{\left\vert Q\right\vert \ell}}{4\pi\ell^{2}}\Delta\phi$, at fixed $Q$.
There are two solitons for each boundary condition for $\Delta\phi< \Delta
\phi_{max}$. The supersymmetric solution is at the maximum value $\Delta\phi=
\Delta\phi_{max}$. }%
\label{figure2}%
\end{figure}

In the Euclidean action, we need to add a boundary term associated with the
change in boundary conditions \cite{Marolf:2006nd}
\begin{equation}
\frac{S_{\partial}}{V}=-\lim_{R\rightarrow\infty}\left(  \frac{1}{2}\sqrt
{h}N_{\mu}A_{\nu}F^{\mu\nu}\right)  _{r=R}=-\frac{2Q^{2}}{\ell^{2}r_{0}},
\end{equation}
so
\begin{equation}
S^{\prime}= S + S_{\partial}= - \left(  \frac{\mu}{2\ell^{2}} + \frac{2Q^{2}%
}{\ell^{2}r_{0}} \right)  V .
\end{equation}
In figure \ref{figure2}, we plot the action for these solutions as a function
of $\Delta\phi$ at fixed $Q$. We see that there are again two branches of
solutions, coalescing at the supersymmetric solution at $\Delta\phi=
\Delta\phi_{max} $. The lower branch corresponds to the solutions with $\mu
<0$, which are always the dominant saddle for these boundary conditions.

\bigskip

\section{Generalization to higher genus}

\label{highg}

It is straightforward to extend this discussion to the Euclidean partition
function of the field theory on $S^{1} \times\Sigma_{g}$, where $\Sigma_{g}$
is a Riemann surface of genus $g >1$, studied in
\cite{Benini:2015noa,Benini:2016hjo,Benini:2015eyy,Bobev:2020pjk}. The
non-linear form of Euclidean gauged $\mathcal{N}=2$ supergravity was
constructed in \cite{deWit:2017cle}. The relevant solutions are obtained from
the analytic continuation of a black hole with horizon $\Sigma_{g}$, carrying
both electric and magnetic charges,
\begin{equation}
ds^{2} = r^{2} ds^{2}_{\Sigma_{g}} + \frac{dr^{2}}{f(r)} + f(r) d\phi^{2},
\end{equation}
with
\begin{equation}
f(r) = \frac{r^{2}}{\ell^{2}} - 1 - \frac{\mu}{r} + \frac{P^{2} - Q^{2}}%
{r^{2}},
\end{equation}
where $ds^{2}_{\Sigma_{g}}$ is the unit constant negative curvature metric on
$\Sigma_{g}$, and we have a gauge field
\begin{equation}
F = 2P \epsilon_{\Sigma_{g}} + \frac{2Q}{r^{2}} d\phi\wedge dr,
\end{equation}
where $\epsilon_{\Sigma_{g}}$ is the volume form of $ds^{2}_{\Sigma_{g}}$.
Note that in the analytic continuation, we have analytically continued the
electric charge $Q$ but not the magnetic charge $P$, hence the difference in
sign in the metric function. We want to restrict the parameters so that $f(r)$
has at least one real positive root; we take as before $r_{0}$ to be the
largest positive root, $f(r_{0}) = 0$.

The solution considered in \cite{Benini:2015eyy} has $\mu=Q=0$, and $P=\pm
\ell/2$. This is a supersymmetric black hole, first studied in
\cite{Romans:1991nq}; the contributions of the field strength and the spin
connection on $\Sigma_{g}$ in \eqref{eq:complex-killing-spinor-eqs} cancel, to
allow us to find two chiral Killing spinors, namely the solution is $1/4$
BPS. For these parameters, $f(r)$ has a double root at $r_{0}=\ell/\sqrt{2}$,
\begin{equation}
f(r)=\left(  \frac{r}{\ell}-\frac{\ell}{2r}\right)  ^{2}.
\end{equation}
The geometry near $r=r_{0}$ is thus $H^{2}\times\Sigma_{g}$; in Lorentzian
signature, this is an extremal black hole, with a near-horizon $AdS_{2}$ region.

The solution is regular for $\phi\in\mathbb{R}$, but we can obtain solutions
with periodic $\phi$ by quotienting this solution, as in our previous
discussion of Poincar\'e-AdS. As before, we can consider adding a Wilson line
along the $\phi$ circle in the quotient; this corresponds to a chemical
potential for the R-charge in the partition function. This quotient will have
unbroken supersymmetry if $\Phi= 2 \pi n \ell$ for integer $n$. Recall that
$\Phi$ is periodic with period $8 \pi\ell$, so there are four physically
inequivalent values giving supersymmetric solutions, which we can take to be
$n=-1,0,1,2$. For odd $n$, the unbroken Killing spinors are antiperiodic on
the $\phi$ circle, while for even $n$, they are periodic.

In \cite{Bobev:2020pjk}, the solutions with $\mu=0$ but non-zero $Q$ were
considered;
\begin{equation}
f(r) = \left(  \frac{r}{\ell} - \frac{\ell}{2 r} \right)  ^{2} - \frac{Q^{2}%
}{r^{2}}%
\end{equation}
with
\begin{equation}
A = A_{\Sigma_{g}}+ \left(  \frac{2Q}{r} - \frac{2Q}{r_{0}} \right)  d \phi,
\end{equation}
where $A_{\Sigma_{g}}$ is a gauge potential on the Riemann surface giving the
magnetic part of $F$. The largest root of $f$ is at
\begin{equation}
r_{0} = \frac{\ell}{\sqrt{2}} \sqrt{1 + |Q|/\sqrt{2} \ell}.
\end{equation}
This is now a single root, so the geometry near $r=r_{0}$ is $\mathbb{R}^{2}
\times\Sigma_{g}$, and we need to choose $\phi$ periodic with period
\begin{equation}
\Delta\phi= \frac{4\pi}{f^{\prime}(r_{0})} = \frac{\pi\ell r_{0}}{|Q|}.
\end{equation}
There is a boundary Wilson line $\Phi= \frac{2Q}{r_{0}} \Delta\phi= \pm
2\pi\ell$, and the solution is supersymmetric, with antiperiodic Killing spinors.

Thus, as in the earlier discussion, there are two supersymmetric solutions
with antiperiodic Killing spinors for $\Phi= \pm2\pi\ell$; the quotiented
solution with $Q=0$ and the soliton-like solution with non-zero $Q$. The
action of these solutions is equal \cite{Bobev:2020pjk},
\begin{equation}
S_{E} = - \frac{\sqrt{2} \pi}{3}(g-1) N^{3/2}.
\end{equation}
This reproduces the value of the twisted topological index calculated in the
field theory using localization methods \cite{Benini:2016hjo}. It is in
principle straightforward to extend this analysis to more general solutions
with scalars, although explicitly constructing the soliton solutions is
challenging \cite{Bobev:2020pjk}.

As in the genus one case, it would be very interesting to understand further
the degeneracy between these two solutions. It would also be interesting to
understand the difference between odd $n$ (where we have this degeneracy) and
even $n$ (where the $Q=0$ solution is the unique supersymmetric solution) from
the field theory perspective.

If we want to consider different values of $\Phi$, we can detune this solution
by considering the geometry with $\mu$ also non-zero,
\begin{equation}
f(r) = \left(  \frac{r}{\ell} - \frac{\ell}{2 r} \right)  ^{2} - \frac{\mu}{r}
- \frac{Q^{2}}{r^{2}}.
\end{equation}
We leave a complete understanding of the space of possible smooth
non-supersymmetric solutions for future work; it is not trivial in this case
to solve for $\mu, Q$ in terms of the natural boundary parameters $\Delta\phi,
\Phi$, or to establish the range of parameters for which there is a positive
root $r_{0}$.


\section{Five dimensions}

\label{fived}

\subsection{Minimal gauged supergravity}

The discussion so far has been in four bulk dimensions; it is easy to
generalise the construction to higher dimensions. Here we will consider in
particular Einstein-Maxwell-AdS in five bulk dimensions constructed in several
papers \cite{Schwarz:1983qr, Howe:1983sra, Schwarz:1983ku, Gunaydin:1984fk,
Kim:1985ez, Pernici:1985ju, Gunaydin:1984qu, Baguet:2015sma}. We follow
\cite{Gauntlett:2003fk, Cvetic:1999xp}. The action principle is
\begin{equation}
S\left(  g,A\right)  =\frac{1}{8\pi G}\int d^{5}x\left(  \sqrt{-g}\left[
\frac{R}{2}+\frac{6}{\ell^{2}}-\frac{1}{8}F_{\mu\nu}F^{\mu\nu}\right]
+\frac{1}{24\sqrt{3}}F\wedge F\wedge A\right)  \text{ }, \label{Lag5d}%
\end{equation}
with field equations%

\begin{align}
&  \partial_{\mu}\big(\sqrt{-g}F^{\mu\nu}\big)=0\,\,,\nonumber\\
&  R_{\mu\nu}-\tfrac{1}{2}g_{\mu\nu}R-\tfrac{1}{2}\big[F_{\mu\rho}\,F_{\nu}%
{}^{\rho}-\tfrac{1}{4}g_{\mu\nu}F_{\rho\sigma}F^{\rho\sigma}\big]-\frac
{6}{\ell^{2}}\,g_{\mu\nu}=0\,.
\end{align}
where we have used that our solutions shall satisfy $F\wedge F=0$. In our
conventions, the Killing spinor equations can be written in terms of a single
complex spinor, $\Psi$, as follows%

\begin{equation}
\big(\partial_{\mu}+\tfrac{1}{4}\omega_{\mu}{\!}^{ab}\gamma_{ab}%
-\frac{\mathrm{i}}{8\sqrt{3}}\gamma_{\mu\lambda\alpha}F^{\lambda\alpha}%
+\frac{\mathrm{i}}{2\sqrt{3}}F_{\mu\alpha}\gamma^{\alpha}\big)\Psi+\ell
^{-1}\left(  \frac{\mathrm{i}}{2}\gamma_{\mu}-\frac{\sqrt{3}}{2}A_{\mu
}\,\right)  \gamma^{4}\Psi^{\ast}=0\text{ .} \label{KSSM}%
\end{equation}
where $\gamma^{4}=\mathrm{i}\gamma^{0}\gamma^{1}\gamma^{2}\gamma^{3}$ is the
gamma matrix along the fifth dimension and $\Psi^{\ast}$ is the complex
conjugate of $\Psi$.

The theory we considered here is obtained by a dimensional reduction of type
IIB\ supergravity over the five-sphere with the ansatz \cite{Cvetic:1999xp}
\begin{align}
ds_{10}^{2}  &  =ds_{5}^{2}+\ell^{2}\sum_{i=1}^{3}d\mu_{i}^{2}+\mu_{i}%
^{2}(d\phi_{i}+\frac{1}{\sqrt{3}\ell}A)^{2}\\
F_{5}  &  =G_{5}+\ast_{5} G_{5}\\
G_{5}  &  =-\frac{4}{\ell}\epsilon_{5}+\frac{\ell^{2}}{\sqrt{3}}\sum_{i=1}%
^{3}\mu_{i}d\mu_{i}\wedge(d\phi_{i}+\frac{1}{\sqrt{3}\ell}A)\wedge\ast_{5} dA
\end{align}
where \thinspace$\ast_{5}$ is the Hodge dual with respect to the
five-dimensional metric $ds_{5}^{2}$ and $\epsilon_{5}$ is its volume form.

\subsection{Soliton solutions}

By double analytic continuation of the electrically charged black hole
solutions with a flat boundary, we obtain a solution
\begin{equation}
ds^{2}=\frac{r^{2}}{\ell^{2}}(-dt^{2}+dy^{2}+dz^{2})+\frac{dr^{2}}%
{f(r)}+f(r)d\phi^{2},
\end{equation}
where
\begin{equation}
f(r)=\frac{r^{2}}{\ell^{2}}-\frac{\mu}{r^{2}}-\frac{q^{2}}{r^{4}}.
\end{equation}
This is a solution of Einstein-Maxwell with the gauge field
\begin{equation}
A=\left(  \frac{\sqrt{3}q}{r^{2}}-\frac{\sqrt{3}q}{r_{0}^{2}}\right)  d\phi,
\end{equation}
where $r_{0}$ is the largest root of $f(r)$, $f(r_{0})=0$. We can write
\begin{equation}
\mu=\frac{(r_{0}^{6}-q^{2}\ell^{2})}{\ell^{2}r_{0}^{2}}.
\end{equation}
The solution is smooth at $r=r_{0}$ if $\phi$ has period
\begin{equation}
\Delta\phi=\frac{4\pi}{f^{\prime}(r_{0})}.
\end{equation}
The flux through the $\phi$ circle at $r\rightarrow\infty$ is $\Phi=\oint
A=\frac{\sqrt{3}q}{r_{0}^{2}}\Delta\phi$. The boundary data is $\Delta\phi$
and $\Phi$, so it is convenient to re-express the bulk parameters in terms of
these; we have
\begin{equation}
q=\frac{r_{0}^{2}}{\sqrt{3}}\frac{\Phi}{\Delta\phi},
\end{equation}
with
\begin{equation}
r_{0}=\frac{\pi\ell^{2}}{2\Delta\phi}\left(  1\pm\sqrt{1-\frac{\Phi^{2}}%
{\Phi_{max}^{2}}}\right)  ,
\end{equation}
where $\Phi_{max}=\sqrt{\frac{3}{2}}\pi\ell$. We see that for $\Phi<\Phi
_{max}$, there are two branches of solutions. The $+$ branch approaches the
AdS soliton as $\Phi\rightarrow0$, while the $-$ branch approaches
Poincar\'{e} - AdS, and they coalesce at $\Phi=\Phi_{max}$. We have $\mu=0$
when $r_{0}^{6}=q^{2}\ell^{2}$, that is when
\begin{equation}
r_{0}^{2}=\frac{\Phi^{2}\ell^{2}}{3\Delta\phi^{2}}\quad\Rightarrow
\quad2-3\frac{\Phi^{2}}{\Phi_{max}^{2}}\pm2\sqrt{1-\frac{\Phi^{2}}{\Phi
_{max}^{2}}}=0,
\end{equation}
which is satisfied on the $+$ branch at $\Phi=\Phi_{S}=\frac{2\sqrt{2}}{3}%
\Phi_{max}$, that is $\Phi_{S}=\frac{2\pi}{\sqrt{3}}\ell$.

\subsection{Supersymmetric solutions}

\bigskip To solve the Killing spinor equation at the supersymmetric point
$\mu=0$, we introduce the change of coordinates
\begin{equation}
r=\left\vert q\ell\right\vert ^{2/3}\cosh(\rho)^{1/3},
\end{equation}

in terms of which the metric and gauge field read%

\begin{align}
ds^{2}  &  =\frac{\ell^{2}}{9}d\rho^{2}+\frac{r_{0}^{2}}{\ell^{2}}\left[
\cosh\left(  \rho\right)  ^{2/3}\left(  -dt^{2}+dy^{2}+dz^{2}\right)
+\frac{\sinh\left(  \rho\right)  ^{2}}{\cosh(\rho)^{4/3}}d\phi^{2}\right]  ,\\
A  &  =\frac{\sqrt{3}q}{r_{0}^{2}}\left(  \frac{1}{\cosh(\rho)^{2/3}%
}-1\right)  d\phi,
\end{align}
where in this section we take $r_{0}=\left\vert q\ell\right\vert ^{1/3}$ and
$\Delta\phi=\frac{2\pi\ell^{5/3}}{3q^{1/3}}$. The calculation shall be carried
in the same basis as in the $D=4$ with the extra vielbein%

\begin{equation}
e^{4}=\frac{r_{0}}{\ell}\cosh(\rho)^{1/3}dy\text{ .}%
\end{equation}

\bigskip There are four independent solutions to the Killing spinor equation
on this background:%
\begin{align}
\Psi_{1} &  =\cosh(\rho)^{-1/3}\left(
\begin{array}
[c]{c}%
s_{\rho}c_{\phi}+\mathrm{i}c_{\rho}s_{\phi}\\
-c_{\rho}c_{\phi}+\mathrm{i}s_{\rho}s_{\phi}\\
c_{\rho}s_{\phi}-\mathrm{i}s_{\rho}c_{\phi}\\
s_{\rho}s_{\phi}+\mathrm{i}c_{\rho}c_{\phi}%
\end{array}
\right)  \text{ },\qquad\Psi_{2}=\cosh(\rho)^{-1/3}\left(
\begin{array}
[c]{c}%
c_{\rho}c_{\phi}+\mathrm{i}s_{\rho}s_{\phi}\\
-s_{\rho}c_{\phi}+\mathrm{i}c_{\rho}s_{\phi}\\
-s_{\rho}s_{\phi}+\mathrm{i}c_{\rho}c_{\phi}\\
-c_{\rho}s_{\phi}-\mathrm{i}s_{\rho}c_{\phi}%
\end{array}
\right)  ,\\
\Psi_{3} &  =\cosh(\rho)^{-1/3}\left(
\begin{array}
[c]{c}%
s_{\rho}s_{\phi}-\mathrm{i}c_{\rho}c_{\phi}\\
-c_{\rho}s_{\phi}-\mathrm{i}s_{\rho}c_{\phi}\\
-c_{\rho}c_{\phi}-\mathrm{i}s_{\rho}s_{\phi}\\
-s_{\rho}c_{\phi}+\mathrm{i}c_{\rho}s_{\phi}%
\end{array}
\right)  \text{ },\qquad\Psi_{4}=\cosh(\rho)^{-1/3}\left(
\begin{array}
[c]{c}%
c_{\rho}s_{\phi}-\mathrm{i}s_{\rho}c_{\phi}\\
-s_{\rho}s_{\phi}-\mathrm{i}c_{\rho}c_{\phi}\\
s_{\rho}c_{\phi}+\mathrm{i}c_{\rho}s_{\phi}\\
c_{\rho}c_{\phi}-\mathrm{i}s_{\rho}s_{\phi}%
\end{array}
\right)  \text{ ,}%
\end{align}
where $s_{\rho}=\sinh\left(  \frac{\rho}{2}\right)  $, $c_{\rho}=\cosh\left(
\frac{\rho}{2}\right)  $, $s_{\phi}=\sin\left(  \frac{\pi}{\Delta\phi}%
\phi\right)  $ and $c_{\phi}=\cos\left(  \frac{\pi}{\Delta\phi}\phi\right)  $,
which implies that the solution is $1/2$ BPS. As a cross-check we verify the
formulae which relate Killing spinors and Killing vectors as given in section
2 of \cite{Gauntlett:2003fk}. The most general Killing spinor for this
solution is%

\begin{equation}
\bigskip\Psi=\sum_{i=1}^{4}c_{i}\Psi_{i}\text{ .}%
\end{equation}
where the $c_{i}$ are real constants. It statisfies that%

\begin{align}
\Psi^{T}\left(  \gamma^{4}\right)  ^{\dagger}\gamma^{0}\Psi &  =0\text{ ,}\\
\Psi^{T}\left(  \gamma^{4}\right)  ^{\dagger}\gamma^{0}\gamma^{\mu}\gamma
^{4}\Psi^{\ast}  &  =K^{\mu}\text{ ,}%
\end{align}
where the components of the Killing vector are%

\begin{align}
K^{t}  &  =\frac{2\ell}{r_{0}}\left(  c_{1}^{2}+c_{2}^{2}+c_{3}^{2}+c_{4}%
^{2}\right)  \text{ ,}\\
K^{\phi}  &  =\frac{4\ell}{r_{0}}\left(  c_{3}c_{4}+c_{1}c_{2}\right)  \text{
,}\\
K^{z}  &  =-\frac{4\ell}{r_{0}}\left(  c_{1}c_{4}-c_{3}c_{2}\right)  \text{
,}\\
K^{y}  &  =-\frac{2\ell}{r_{0}}\left(  c_{1}^{2}-c_{2}^{2}+c_{3}^{2}-c_{4}%
^{2}\right)  \text{ .}%
\end{align}

Let us note that in an asymptotic expansion, the leading term of the spinors is%

\begin{align}
\lim_{\rho\rightarrow\infty}e^{-\frac{\rho}{6}}\Psi_{1}  &  =\left(
\begin{array}
[c]{c}%
\exp\left(  \mathrm{i}\frac{\pi}{\Delta\phi}\phi\right) \\
-\exp\left(  -\mathrm{i}\frac{\pi}{\Delta\phi}\phi\right) \\
-\mathrm{i}\exp\left(  \mathrm{i}\frac{\pi}{\Delta\phi}\phi\right) \\
\mathrm{i}\exp\left(  -\mathrm{i}\frac{\pi}{\Delta\phi}\phi\right)
\end{array}
\right)  ,\qquad\lim_{\rho\rightarrow\infty}e^{-\frac{\rho}{6}}\Psi
_{2}=\left(
\begin{array}
[c]{c}%
\exp\left(  \mathrm{i}\frac{\pi}{\Delta\phi}\phi\right) \\
-\exp\left(  -\mathrm{i}\frac{\pi}{\Delta\phi}\phi\right) \\
\mathrm{i}\exp\left(  \mathrm{i}\frac{\pi}{\Delta\phi}\phi\right) \\
-\mathrm{i}\exp\left(  -\mathrm{i}\frac{\pi}{\Delta\phi}\phi\right)
\end{array}
\right)  ,\\
\lim_{\rho\rightarrow\infty}e^{-\frac{\rho}{6}}\Psi_{3}  &  =\left(
\begin{array}
[c]{c}%
-\mathrm{i}\exp\left(  \mathrm{i}\frac{\pi}{\Delta\phi}\phi\right) \\
-\mathrm{i}\exp\left(  -\mathrm{i}\frac{\pi}{\Delta\phi}\phi\right) \\
-\exp\left(  \mathrm{i}\frac{\pi}{\Delta\phi}\phi\right) \\
-\exp\left(  -\mathrm{i}\frac{\pi}{\Delta\phi}\phi\right)
\end{array}
\right)  ,\qquad\lim_{\rho\rightarrow\infty}e^{-\frac{\rho}{6}}\Psi
_{4}=\left(
\begin{array}
[c]{c}%
-\mathrm{i}\exp\left(  \mathrm{i}\frac{\pi}{\Delta\phi}\phi\right) \\
-\mathrm{i}\exp\left(  -\mathrm{i}\frac{\pi}{\Delta\phi}\phi\right) \\
\exp\left(  \mathrm{i}\frac{\pi}{\Delta\phi}\phi\right) \\
\exp\left(  -\mathrm{i}\frac{\pi}{\Delta\phi}\phi\right)
\end{array}
\right)  \text{.}%
\end{align}

On pure AdS$_{5}$, the theory has four Killing spinors; if we introduce a
Wilson loop on this background with $A=A_{\phi}d\phi$. Indeed, a basis for the
local solutions to the Killing spinor equation in the Poincar\'{e} patch are%

\begin{equation}
\Psi_{1+}^{AdS}=\Delta_{+}\left(
\begin{array}
[c]{c}%
0\\
1\\
0\\
0
\end{array}
\right)  ,\text{ }\Psi_{1-}^{AdS}=\Delta_{-}\left(
\begin{array}
[c]{c}%
-1\\
0\\
0\\
0
\end{array}
\right)  ,\text{ }\Psi_{2+}^{AdS}=\Delta_{+}\left(
\begin{array}
[c]{c}%
0\\
0\\
0\\
1
\end{array}
\right)  ,\text{ }\Psi_{2-}^{AdS}=\Delta_{-}\left(
\begin{array}
[c]{c}%
0\\
0\\
1\\
0
\end{array}
\right)  \text{,}%
\end{equation}

\begin{align}
\Psi_{3+}^{AdS}  &  =\Delta_{+}\left(
\begin{array}
[c]{c}%
0\\
\phi+t\\
\ell^{2}r^{-1}\\
z+\mathrm{i}y
\end{array}
\right)  ,\text{ }\Psi_{3-}^{AdS}=\Delta_{-}\left(
\begin{array}
[c]{c}%
-\phi-t\\
0\\
z-\mathrm{i}y\\
-\ell^{2}r^{-1}%
\end{array}
\right)  ,\\
\Psi_{4+}^{AdS}  &  =\Delta_{+}\left(
\begin{array}
[c]{c}%
\ell^{2}r^{-1}\\
z-\mathrm{i}y\\
0\\
t-\phi
\end{array}
\right)  ,\Psi_{4-}^{AdS}=\Delta_{-}\left(
\begin{array}
[c]{c}%
-z-\mathrm{i}y\\
\ell^{2}r^{-1}\\
-\phi+t\\
0
\end{array}
\right)  \text{ }.
\end{align}

where $\Delta_{\pm}=\sqrt{r}\exp\left(  \pm\mathrm{i}\frac{\sqrt{3}A_{\phi}%
}{2\ell}\phi\right)  $. The most general solution to the Killing spinor
equation in this basis is%
\begin{equation}
\Psi^{AdS}=\sum_{i=1}^{4}\alpha_{i}\Psi_{i+}^{AdS}+\sum_{i=1}^{4}\alpha
_{i}^{\ast}\Psi_{i-}^{AdS}%
\end{equation}
where the constants $\alpha_{i}^{\ast}$ are the complex conjugate of
$\alpha_{i}$. Again we can see from the Killing spinors that the
identification in $\phi$ breaks half of the supersymmetries of AdS$_{5}$.

For $\Phi=\Phi_{S}=\frac{2\pi}{\sqrt{3}}\ell$, we can also obtain a
supersymmetric solution from Poincar\'{e}-AdS with a constant Wilson loop
along the $\phi$ circle in the bulk. As in four dimensions, this is easily
seen from the uplift of the theory to ten dimensions. The theory we considered
here is obtained by a dimensional reduction \cite{Cvetic:1999xp}
\begin{equation}
ds_{10}^{2}=ds_{5}^{2}+\ell^{2}\sum_{i=1}^{3}d\mu_{i}^{2}+\mu_{i}^{2}%
(d\phi_{i}+\frac{1}{\sqrt{3}\ell}A)^{2};
\end{equation}
The coordinates $\phi_{i}$ are $2\pi$ periodic. The ten-dimensional metric is
thus invariant under shifts of the Wilson loop by $\Phi\rightarrow\Phi
+2\pi\sqrt{3}\ell$, making this a periodic variable. Quotienting
Poincar\'{e}-AdS with a Wilson loop on the $\phi$ circle corresponds to a
quotient of $AdS_{5}\times S^{5}$ by $(\phi,\tilde{\phi}_{i})\sim(\phi
+\Delta\phi,\tilde{\phi}_{i}+\Phi/\sqrt{3}\ell)$, so the ten-dimensional
Killing spinors will be invariant under the quotient if $\Phi=\frac{2\pi
}{\sqrt{3}}n\ell$ for integer $n$. Given the periodicity of $\Phi$, there are
three physically inequivalent cases, $n=-1,0,1$. For $n=-1,1$, we have a
degeneracy between the Poincar\'{e}-AdS solution and the supersymmetric
soliton. The main qualitative difference between the four- and
five-dimensional cases is that we only have one case with a unique
supersymmetric solution here.

\section{Conclusions}

We have shown that the simplest $D=4$ and $D=5$ gauged supergravities
theories, which allow for charged solutions, contain in its spectrum of
supersymmetric solutions Lorentzian planar solitons. These solitons can be
seen as a supersymmetric extension of the rather well-known AdS soliton of
Horowitz and Myers \cite{Horowitz:1998ha}. The existence of these
supersymmetric solitons poses an holographic puzzle.  One usually expects
that for given boundary conditions, at least with sufficient supersymmetry,
the gravitational solution would be completely fixed. This solution would then
be dual to a supersymmetric state in the dual CFT. However, as we have
explictly shown this is not the case. One possible way out would be if the solutions were supersymmetric with respect to different realizations of the
supersymmetry algebra in $AdS_{4}$ \cite{Hristov:2011ye}. However in the different realizations, for the same
supersymmetry and same boundary conditions for the metric, the form of the
Killing spinors is different. Our explicit construction of the
Killing spinors shows that the soliton Killing spinors are asymptotically the
Killing spinors of a locally AdS spacetime. Hence, these configurations are
indeed completely indistiguishable from the boundary point of view. Moreovoer,
the Killing spinors for the solitons provide an smooth interpolation between
some Killing spinors of Minkowski$_{D}$ and AdS$_{D}$, for $D=4$ and $D=5$.

The simplicity of the susy AdS soliton make us conjecture that these $1/2$ BPS
solutions can be generalized to have different charges by the introduction of
running scalars in $D=4$ and $D=5$ gauged supergravity \cite{Cvetic:1999xp,
Anabalon:2019tcy, Anabalon:2020qux, Anabalon:2020pez}. Furthermore, we believe
that this susy AdS soliton should also be part of the spectrum of any gauged
supergravity in any dimension that contains a consistent supersymmetric
truncation with a $U(1)$ gauge field. Therefore the degeneracy of susy states
we presented here, it is very likely, a generic feature of these boundary conditions.

The structure of the soliton seems to be well suited to be interpreted in the
context of holographic superconductivity \cite{Hartnoll:2008vx}. Indeed, from
the holographic point of view, the boundary of the soliton is a cylinider with
a current flowing along the circle. The current is flowing without any
electromagnetic field on the surface of the cylinder and there is a magnetic
field piercing the cylinder. It would be worth understanding if the periodic
oscillation of the critical temperature of the superconductor takes place
holographically as expected due to the Little-Parks effect \cite{LP}.

An interesting question to be further explored is whether other boundary
conditions can yield a degeneration in the spectrum of supersymmetric
solutions. Asymptotically locally AdS, everywhere regular supersymmetric solutions in the pure
Einstein-Maxwell theory with a negative cosmological constant exist in four
dimensions but with a much more complex structure than the one discussed here
\cite{Anabalon:2020loe}.

\section*{Acknowledgements}

We thank Bernard de Wit, Jerome Gauntlett and Mario Trigiante for enlightening
discussions and comments. We would like to thank the support of Proyecto de
cooperaci\'{o}n internacional 2019/13231-7 FAPESP/ANID. The research of AA is
supported in part by the Fondecyt Grants 1210635, 1181047 and 1200986. SFR is
supported in part by STFC through grant ST/T000708/1.


\bigskip

\bigskip

\bibliographystyle{utphys}
\bibliography{solitons}

\end{document}